\documentclass[twocolumn,aps,showpacs,floats,letterpaper,floatfix,groupedaddress,eqsecnum]{revtex4}

\usepackage{amssymb,amsmath}
\usepackage[dvips]{graphicx}

\usepackage{dcolumn,epsfig}

\newcommand{\beq}{\begin{equation}}
\newcommand{\eeq}{\end{equation}}
\newcommand{\bes}{\begin{subequations}}
\newcommand{\ees}{\end{subequations}}
\newcommand{\bea}{\begin{eqnarray}}
\newcommand{\eea}{\end{eqnarray}}
\newcommand{\ba}{\begin{array}}
\newcommand{\ea}{\end{array}}
\newcommand{\beqn}{\begin{eqnarray*}}
\newcommand{\eeqn}{\end{eqnarray*}}

\newcommand{\f}[2]{\frac{#1}{#2}}

\newcommand{\n}{\eta}

\newcommand{\la}{\langle}
\newcommand{\dg}{\dagger}
\newcommand{\ra}{\rangle}
\def\bk{{\bf k}}

\def\bq{{\bf q}}

\def\bx{{\bf x}}
\def\b0{{\bf 0}}

\def\nn{\nonumber}

\newlength{\sizeonefig}
\newlength{\sizetwofig}
\begin{document}
\title{Correlated few-photon transport in one-dimensional waveguides: linear and nonlinear dispersions}

\author{Dibyendu Roy$^\dg$} 
\affiliation{Department of Physics, University of California-San Diego, La Jolla, California 92093-0319, USA}

\begin{abstract}
We address correlated few-photon transport in one-dimensional waveguides coupled to a two-level system (TLS), such as an atom or a quantum dot.  We derive exactly the single-photon and two-photon current (transmission) for linear and nonlinear (tight-binding sinusoidal) energy-momentum dispersion relations of photons in the waveguides and compare the results for the different dispersions. A large enhancement of the two-photon current for the sinusoidal dispersion has been seen at a certain transition energy of the TLS away from the single-photon resonances. 
\end{abstract}



\maketitle
\section{Introduction}
 The study of equilibrium dynamics of correlated photons and mixed photonic-atomic excitations (called polaritons) has received huge interest in the last few years. It allows one to simulate strongly correlated condensed matter phenomena, for example, quantum phase transition in simple controllable quantum systems \cite{Greentree06}, such as coupled optical cavities with each containing a single two-level system (TLS)\cite{Angelakis07}. It is especially interesting to model nonequilibrium dynamics in these systems. Now, it is a highly nontrivial task to study out-of-equilibrium dynamics of many correlated  photons nonperturbatively. Instead, we pose a relatively simpler problem: we examine the exact dynamics of  few correlated  photons in one dimension. 

The dynamics of few correlated photons in the Dicke-type Hamiltonian with a tight-binding (TB) sinusoidal energy dispersion relation for photons was investigated recently using a computational technique \cite{Longo10}. The single-photon dynamics in that model was studied before in Ref.\cite{Zhou08} to demonstrate a quantum switch for the coherent transport of a single photon. The two-photon scattering states in a similar model for a linear energy dispersion relation of photons was derived exactly by employing a generalized Bethe ansatz method \cite{Shen07} and Lehmann-Symanzik-Zimmermann reduction techniques \cite{Shi09}. Recently large-scale ultrahigh-$Q$ coupled resonator arrays with a TB sinusoidal dispersion relation have been realized in photonic crystals \cite{Notomi08}. 

Here we study correlated few-photon transport for a TB sinusoidal dispersion relation with band edges using an analytical method \cite{Dhar08, Roy09}. We then compare the results for the nonlinear sinusoidal dispersion relation with a linear dispersion relation. We apply the Bethe ansatz method to study the linear dispersion \cite{Shen07, Nishino09, Roy10}. There are some studies on correlated photon transport for cosine dispersion \cite{Shi09}, but none has tried before to compare correlated dynamics in the waveguides for different dispersions. Shi, Fan, and Sun \cite{Shi10} considered correlated two-photon transport in a linear waveguide coupled to cavity with a TLS to study a photon blockade in the strong TLS-cavity coupling regime \cite{Birnbaum05}. It is possible to further extend the method of this paper to investigate correlated photon dynamics in a cavity coupled to a nonlinear waveguide.

\section{Model}
We consider a system consisting of two one-dimensional (1D) coupled-resonator-optical waveguides  being connected by a TLS, such as  an atom or a quantum dot \cite{Zhou08}. The two-photon or multiphoton dynamics in this system is strongly correlated \cite{Longo10, Shen07, Roy10}. The full Hamiltonian within the random-phase approximation is given by 
\bea
\mathcal{H}_n&=&-J\sum_{x=-\infty}^{\infty}\hspace{-0.1cm}'(a_x^{\dg}a_{x+1}+a_{x+1}^{\dg}a_x)+\hbar\Omega\sigma_z/2\nn\\&+&V_L(a_{-1}^{\dg}\sigma_{-}+\sigma_{+}a_{-1})+ V_R(a_{1}^{\dg}\sigma_{-}+\sigma_{+}a_{1}).
\eea 
Here, $a_x^{\dg}$ $(a_x)$ denotes the photon creation (annihilation) operator at site $x$, and $J$ is the strength of hopping between nearest-neighbor sites. $\sum'$ indicates summation over all integers, omitting $x=-1,0$. The Pauli operators $\sigma_z$ and $\sigma_{\pm}~(\equiv \sigma_x\pm i\sigma_y)$ represent the TLS at the zeroth site. The TLS with a transition frequency $\Omega$ is coupled to the left and  right waveguides by $V_L$ and $V_R$, respectively. Here we assume a direct coupling between the waveguides and the TLS. A generalization of the present results for a sidecoupling is straightforward. The energy dispersion of photons in the waveguides is given by $E_k=-2J\cos k$ with a wave number $-\pi<k<\pi$. Henceforth, subscript or  superscript $n$ and $l$ denote properties pertaining to nonlinear and linear dispersions, respectively. 

Now, we redefine  $\mathcal{H}_n$ in terms of a new Hamiltonian $\mathcal{H}$, where we replace the TLS by an appropriate additional bosonic system at the zeroth site \cite{Leggett87, Longo10}. Then $\mathcal{H}=\mathcal{H}_0+\mathcal{V}$, with
\bea
\mathcal{H}_0&=&-J\sum_{x=-\infty}^{\infty}\hspace{-0.15cm}' ~~(a_x^{\dg}a_{x+1}+a_{x+1}^{\dg}a_x)+\hbar\Omega b^{\dg}b\nn \\
&+&V_L(a_{-1}^{\dg}b+b^{\dg}a_{-1})+ V_R(a_{1}^{\dg}b+b^{\dg}a_{1})~,\nn\\
\mathcal{V}&=&\f{U}{2}~b^{\dg}b(b^{\dg}b-1)~,\label{Ham2}
\eea
where $b$ is the photon annihilation operator at the zeroth site. The ground and excited states of the TLS in $\mathcal{H}_n$ correspond to no photons and a single photon, respectively, at the zeroth site of $\mathcal{H}$. The unwanted multiphoton occupancies at the TLS have been avoided by introducing the interaction term $\mathcal{V}$. We expect that the Hamiltonian in Eq.(\ref{Ham2}) in the limit $U\to \infty$ is equivalent to the Hamiltonian $\mathcal{H}_n$. We define the current operator, $\hat I_x~=~ -i ~V_x (a^{\dg}_x a_{x+1} - a^{\dg}_{x+1} a_x)$, using a continuity equation, where $V_x=-V_L~(-V_R)$ for $x=-1~(1)$, and $V_x=J$ for all other $x$.

A linear energy dispersion for photons, i.e., $E_k=v_gk$ is a good approximation in many situations for 1D optical waveguides, such as microwave transmission lines \cite{Astafiev10} and surface plasmon modes of a metallic nanowires \cite{Akimov07}. Here, $v_g$ is the momentum-independent group velocity of photons. The real-space Hamiltonian of the 1D waveguides and TLS  for a linear energy dispersion is given by
\bea
&&\mathcal{H}_l=-iv_g \sum_{\alpha=1,2}\int dx ~c^{\dagger}_{\alpha}(x)\f{\partial}{\partial x}c_{\alpha}(x)+\hbar\Omega b^{\dagger}b \nn \\
&&+(V_1c_1^{\dg}(0)b+V_2c_2^{\dg}(0)b+H.c.)+\f{U}{2}b^{\dagger}b(b^{\dagger}b-1)~,~
\label{Hamlin}
\eea
where $c^{\dg}_{1}(x)~[c^{\dg}_{2}(x)]$ is a bosonic operator creating a photon at $x$ in the left-hand [right-hand] side of the impurity.  We define the current operator in the system as $\hat{I}=-i[\mathcal{H}_l, N_1-N_2]/2$, where $N_1~[N_2]$ is the total number of photons in the left-hand [right-hand] side of the TLS. Then we derive $\hat{I}=i[V_1c^{\dg}_1(0)b-V_2c^{\dg}_2(0)b-H.c.]/2$. One can readily check that both current operators $\hat I_x$ and $\hat I$ are equivalent. We need to derive exact scattering eigenstates of $\mathcal{H}_l$ to find an average steady-state current in the system. Employing a standard transformation to even-odd modes, $c_{1}(x)=[V_1c_{e}(x)+V_2c_{o}(x)]/V$ and $c_{2}(x)=[V_2c_{e}(x)-V_1c_{o}(x)]/V$ with $V=\sqrt{V_1^2+V_2^2}$, the Hamiltonian in Eq.(\ref{Hamlin}) breaks into two decoupled parts, i.e., $\mathcal{H}_l=\mathcal{H}_e+\mathcal{H}_o$, where 
\bea
\mathcal{H}_e &=&-i v_g\int dx~ c^\dg_{e}(x)\f{\partial}{\partial x}c_{e}(x) + \hbar\Omega b^{\dagger}b\nn\\&+& \f{U}{2}b^{\dagger}b(b^{\dagger}b-1)+ V \big(c^\dg_{e}(0)b+b^{\dg}c_{e}(0)\big),~{\rm and}\nn\\\mathcal{H}_o &=&-i v_g\int dx~ c^\dg_{o}(x)\f{\partial}{\partial x}c_{o}(x).
\eea

\section{Single-photon dynamics}
\begin{figure}
\begin{center}$
\begin{array}{cc}
\includegraphics[width=1.5in]{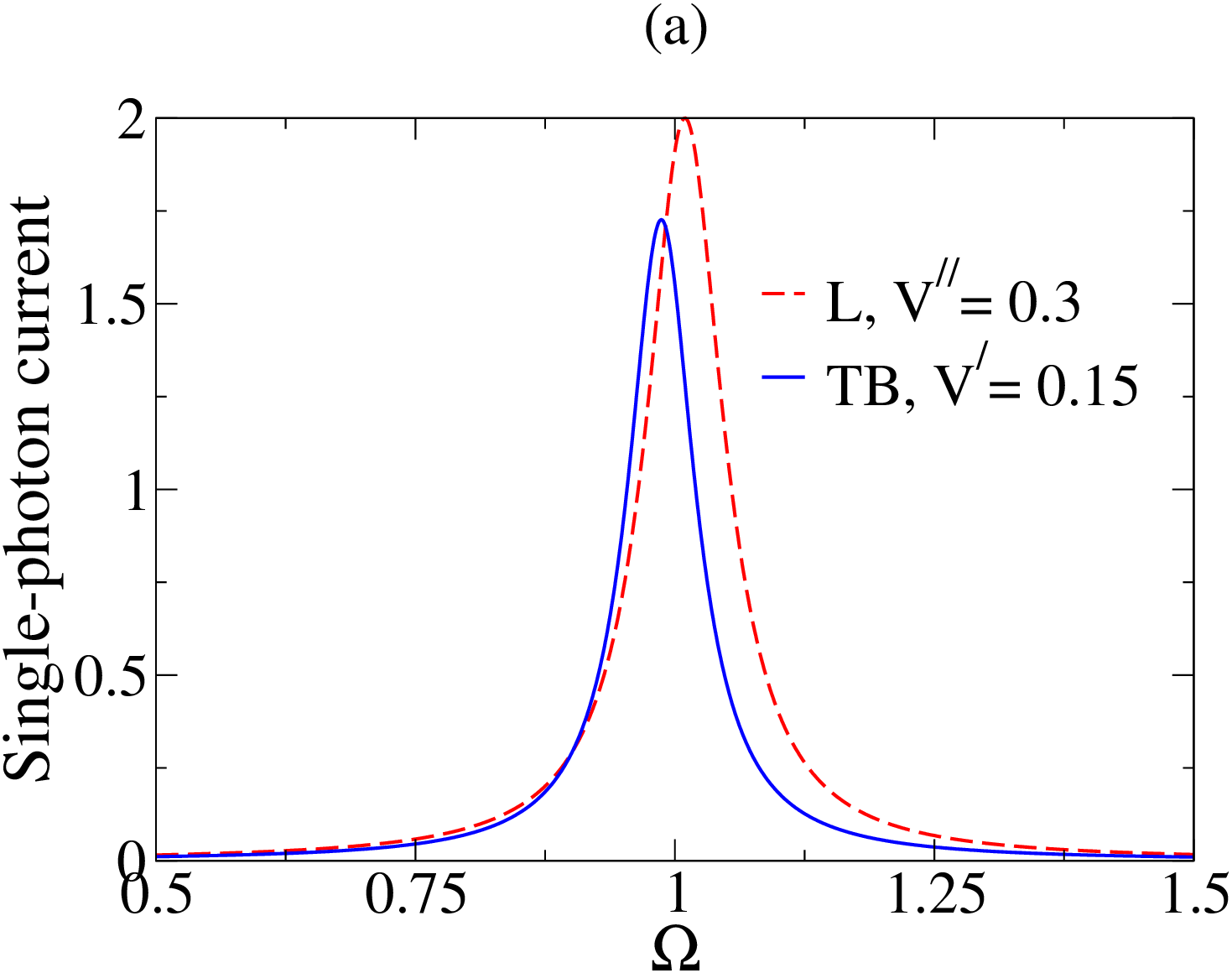} &
\includegraphics[width=1.5in]{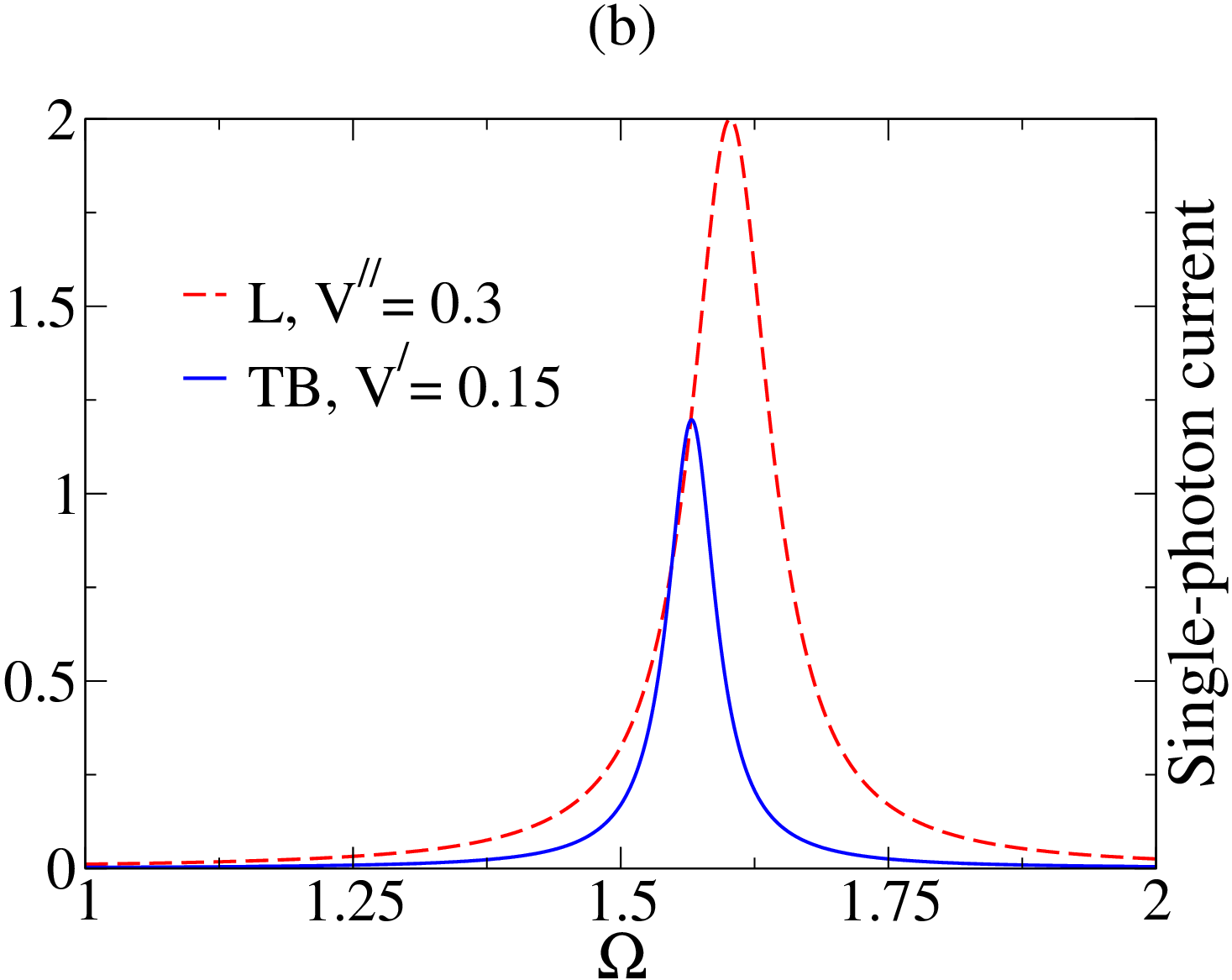}
\end{array}$
\end{center}
\caption{(Color online) Plot of the single-photon current ($I^n(k), I^l(k)$) with $\Omega$ for a linear (L) and a tight-binding (TB) energy dispersion with incident photon energy (a) $E_{k}=1.01$ and (b) $E_{k}=1.6$. Here, $E_{k}$, $\Omega$, $V'$, and $V''$ are in units of $J$, with $J=1$ and $\hbar=1$.}
\label{pl1}
\end{figure}

It is simple to find the single-photon scattering state for the Hamiltonian in Eq.(\ref{Ham2}). Let us denote the single-photon scattering state by $\phi_{k}(x)~(\equiv \la x|\phi_k\ra)$ at lattice site $x$ with incident momentum $k$.
For a photon being incident from the left (a right-moving photon), i.e.,  $0<k<\pi$, we find $\phi_k(x)=e^{ikx}+r^n_ke^{-ikx}$ for $x<0$, $\phi_k(x)=t^n_ke^{ikx}$ for $x>0$, and $\phi_k(0)=-J(1+r^n_k)/V_L$ with
\bea
t^n_k=\f{2iV_LV_R \sin k}{e^{ik}(V_L^2+V_R^2)-J(\hbar\Omega-E_k)},~r^n_k=\f{V_L}{V_R}t^n_k-1~.\label{trans1}
\eea 
The transmission coefficient, $T^n_k=|t^n_k|^2$ shows a Breit-Wigner-like (i.e., Lorentzian) line shape around the resonance transition energy $\hbar\Omega=E_k(2J^2-V_L^2-V_R^2)/2J^2$. For symmetric coupling, i.e., $V_L=V_R=V'$, a single photon at the single-photon resonance can fully transmit from one side of the impurity to the other side. The single-photon current for the nonlinear sinusoidal dispersion is given by $I^n(k)=\la \phi_k|\hat{I}_x|\phi_k\ra=2JT^n_k\sin k$, where $2J\sin k$ is the velocity of a photon. We plot $I^n(k)$ for different $V'$ and $E_k$ in Fig.\ref{pl1}.  

The single-photon scattering state $|1,k\ra$ of $\mathcal{H}_l$ for an incoming photon from the left with momentum $k$ is given by
\bea
\int \f{dx}{\sqrt{2\pi}}\{\f{V_1}{V}[g_k(x)c^{\dg}_e(x)+\delta(x)e_kb^{\dg}]+\f{V_2}{V}h_k(x)c^{\dg}_{o}(x)\}|0,0\ra,\nn 
\eea
where $g_k(x)=e^{ikx}\big[\theta(-x)+\tau_k\theta(x)\big],~h_k(x)=e^{ikx},e_k=V/(v_gk-\hbar\Omega+iV^2/2v_g),~\tau_k=e_k/e_k^*$, with $\theta(x)$ being the step function.  $|n,m\ra$ denotes $n$ photons in the waveguide and $m$ photons at the impurity site (TLS).  We determine the transmission coefficient $T^l_k$ for a photon from the left to the right lead by rearranging the single-photon state  in terms of the original field operators. It is given by  
\bea
T^l_k=\f{4 \Gamma_1\Gamma_2}{(v_gk-\hbar\Omega)^2+(\Gamma_1+\Gamma_2)^2}\label{transl}
\eea
with $\Gamma_{\alpha}=V_{\alpha}^2/2v_g$. The single-photon reflection coefficient is $R^l_k=(V_1^4+V_2^4)/V^4+2V_1^2V_2^2 {\rm Re}[\tau_k]/V^4$.

We can also compute $T^l_k$ from Eq.(\ref{trans1}) after linearizing the energy dispersion along with a proper regularization scheme. This occurs at the matching condition for the wavevector $lk \sim \pi/2$; thus, we find $E_k \sim \pm 2Jk-J\pi$ by expanding the cosine around $k=\pm \pi/2$ and $v_g=2J$ (and lattice constant $l=1$ here). We further require renormalization of the coupling as $V_L=V_1/2$ and $V_R=V_2/2$ in Eq.(\ref{trans1}). The steady-state single-photon current for the linear dispersion is given by $I^l(k)=\la 1,k|\hat{I}|1,k\ra=v_gT^l_k/2\pi$. We need to multiply $I^l(k)$ by $2\pi$ to compare $I^l(k)$ with $I^n(k)$. This is due to our convention for the normalization of the scattering states. The current  $I^l(k)$ also shows a Breit-Wigner-type line shape around the resonance $\hbar\Omega=v_gk$. We plot $I^l(k)$ (after multiplying by $2\pi$) for different $V_1=V_2=V''$ and $E_k$ in Fig. \ref{pl1} and compare it with that for the TB sinusoidal dispersion. We find that the line-shapes of the single-photon current are qualitatively similar for  both linear and  TB sinusoidal dispersions. The differences in the width and the height of the current line shapes for the two dispersions are due to the variance in the group velocity, which is strongly $k$-dependent for the sinusoidal dispersion. 

\section{Two-photon dynamics}
\begin{figure}
\includegraphics[width=7.0cm]{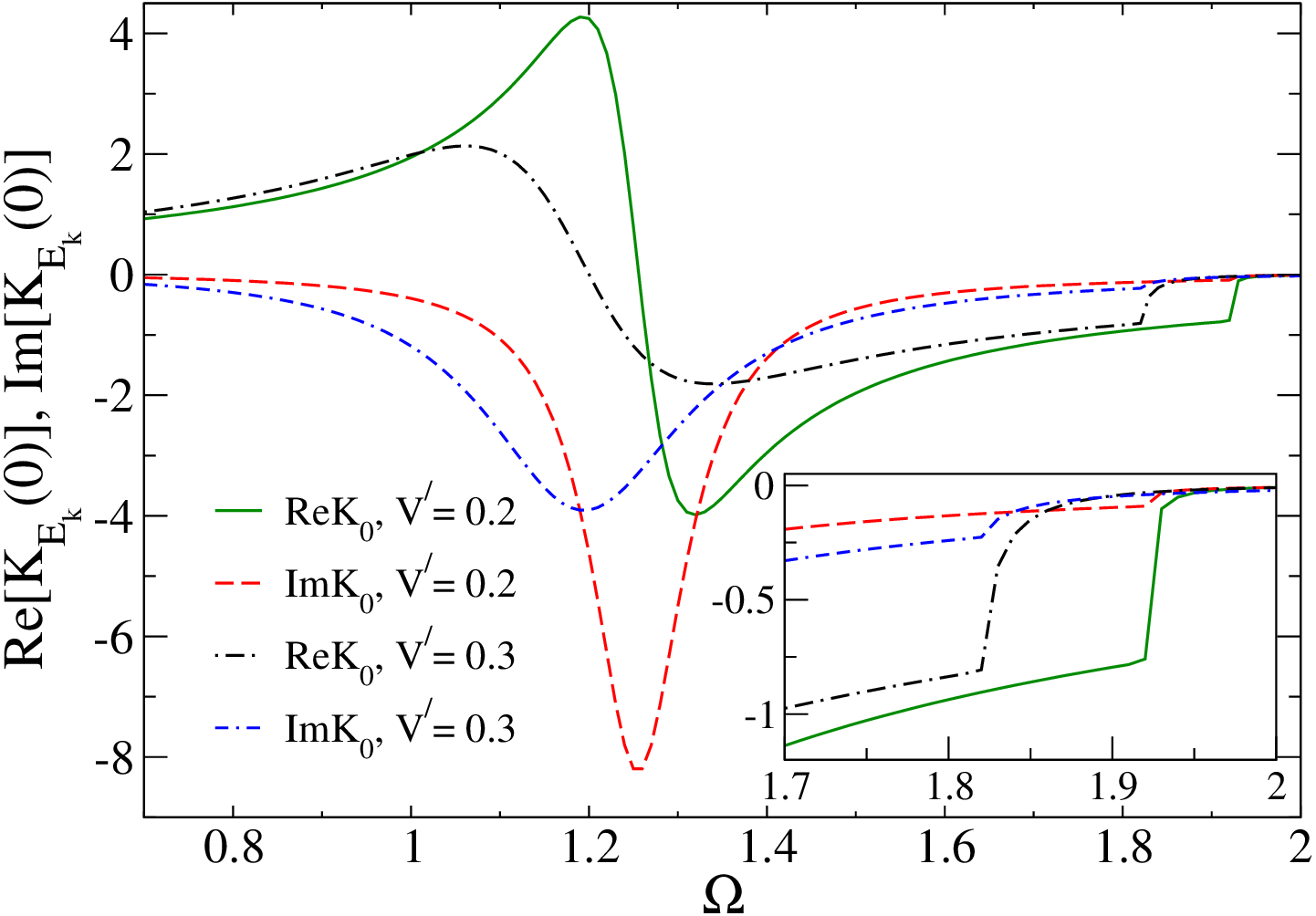}
\caption{(Color online) Plot of the real (Re$K_0$) and the imaginary (Im$K_0$) parts of $K_{E_\bk}(\b0)$ for $E_{k_1}=1.01,E_{k_2}=1.6$, and $V'=0.2,0.3$. Inset shows a magnification of the rightmost part of the plot. $E_{k_1}$, $E_{k_2}$, $\Omega$, and $V'$  are in units of $J$, with $J=1$ and $\hbar=1$.}
\label{pl2} 
\end{figure}
\begin{figure}
\includegraphics[width=7.0cm]{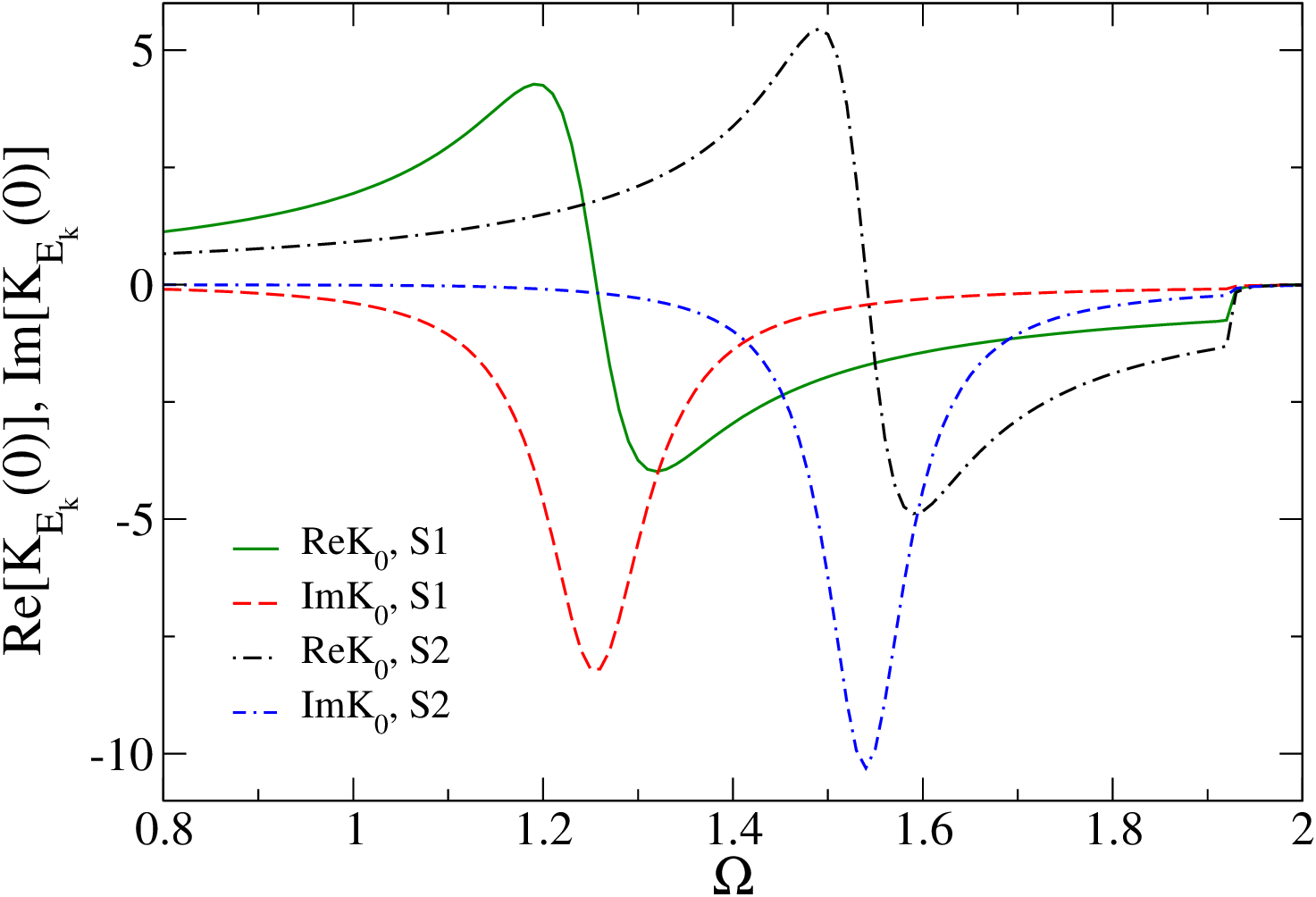}
\caption{(Color online) Plot of the real (Re$K_0$) and the imaginary (Im$K_0$) parts of $K_{E_\bk}(\b0)$ for two sets of parameters. S1$\equiv\{E_{k_1}=1.01,E_{k_2}=1.6\}$, S2$\equiv\{E_{k_1}=1.6,E_{k_2}=1.6\}$ with $V'=0.2$ in both sets. $E_{k_1}$, $E_{k_2}$, $\Omega$, and $V'$ are in units of $J$, with $J=1$ and $\hbar=1$.}
\label{pl3} 
\end{figure}

{\it Tight-binding sinusoidal dispersion:} We study two-photon transport in the lattice model using a technique based on the Lippman-Schwinger scattering theory. This technique was developed recently to discuss electron transport in quantum impurity models \cite{Dhar08, Roy09}. We here assume that the zeroth site is initially empty. The two-photon scattering states $\Psi_{\bk}(\bx)$ of the Hamiltonian $\mathcal{H}$ for arbitrary value of $U$ are given by (see Appendix \ref{app1})
\bea
\Psi_{\bk}(\bx) \equiv \la \bx|\Psi_{\bk}\ra=\phi_{\bk}(\bx)+\f{K_{E_{\bk}}(\bx)\phi_{\bk}({\b0})}{1/U-K_{E_{\bk}}({\b0})}~,\label{2phscatts}
\eea
where $\bx \equiv (x_1,x_2)$, $\bk \equiv (k_1,k_2)$, $\b0 \equiv (0,0)$, the symmetrized initial state $\phi_{\bk}(\bx)=[\phi_{k_1}(x_1)\phi_{k_2}(x_2)+\phi_{k_1}(x_2)\phi_{k_2}(x_1)]/\sqrt{2}$, and the total energy of the two incident photons is $E_{\bk}=-2J(\cos k_1+\cos k_2)$. The expression $K_{E_\bk}(\bx) \equiv \la \bx | G_0^+(E_\bk) |\b0 \ra$ has the explicit form 
\bea
 K_{E_\bk}(\bx) ~&=&~ \f{1}{2} \int_{-\pi}^\pi \int_{-\pi}^\pi \f{dq_1 
dq_2}{(2 \pi)^2} \f{\phi_{\bq} (\bx) \phi_{\bq}^* (\b0)}{E_\bk - E_\bq + i\n}~.\nn
\eea
We find the two-photon scattering state $\psi_{\bk}(\bx)$ of the Hamiltonian $\mathcal{H}_n$ by taking the limit $U\to \infty$ in Eq. (\ref{2phscatts}): 
\bea
\psi_{\bk}(\bx)=\la \bx|\psi_{\bk}\ra=\phi_{\bk}(\bx)-\f{K_{E_{\bk}}(\bx)\phi_{\bk}({\b0})}{K_{E_{\bk}}({\b0})}~.\label{2phscatts2}
\eea
We can immediately see from Eq.(\ref{2phscatts2}) that $\psi_{\bk}(\b0)=0$; i.e., both photons cannot be at the TLS (or zeroth site) simultaneously, but the occupation probability of a single photon at the zeroth site has a finite value. A single photon at the zeroth site corresponds to the excited state of the TLS. The average occupation of the impurity site is calculated by taking the expectation value of  the operator $n_b~(\equiv b^{\dg}b)$ in $|\psi_{\bk}\ra$. $\la n_b \ra=\sum_{x_1}|\la x_1,0|\psi_{\bk}\ra|^2$. The nature of the scattering states $\psi_{\bk}(\bx)$ depends on $K_{E_{\bk}}({\b0})$. Interestingly, we find that both the real and the imaginary parts of $K_{E_{\bk}}({\b0})$ become zero for a broad range of the transition energy $\hbar\Omega$ away from any of the single-photon resonances; for example, $K_{E_{\bk}}({\b0})$ jumps to zero near $\Omega=1.92$ for $V'=0.2$ as shown in Fig. \ref{pl2}.  We plot the real and the imaginary part of $K_{E_{\bk}}({\b0})$ in Figs. \ref{pl2} and \ref{pl3} for different incident energy and coupling strength. The zero of $K_{E_{\bk}}({\b0})$ depends on the strength of coupling $V'$ (see Fig. \ref{pl2}) but is independent of the incident energy, as shown in Fig. \ref{pl3}. We also  notice that the real part of $K_{E_{\bk}}({\b0})$ passes through zero between two single-photon resonance energies.

\begin{figure*}
\centering
\begin{tabular}{ccc}
\epsfig{file=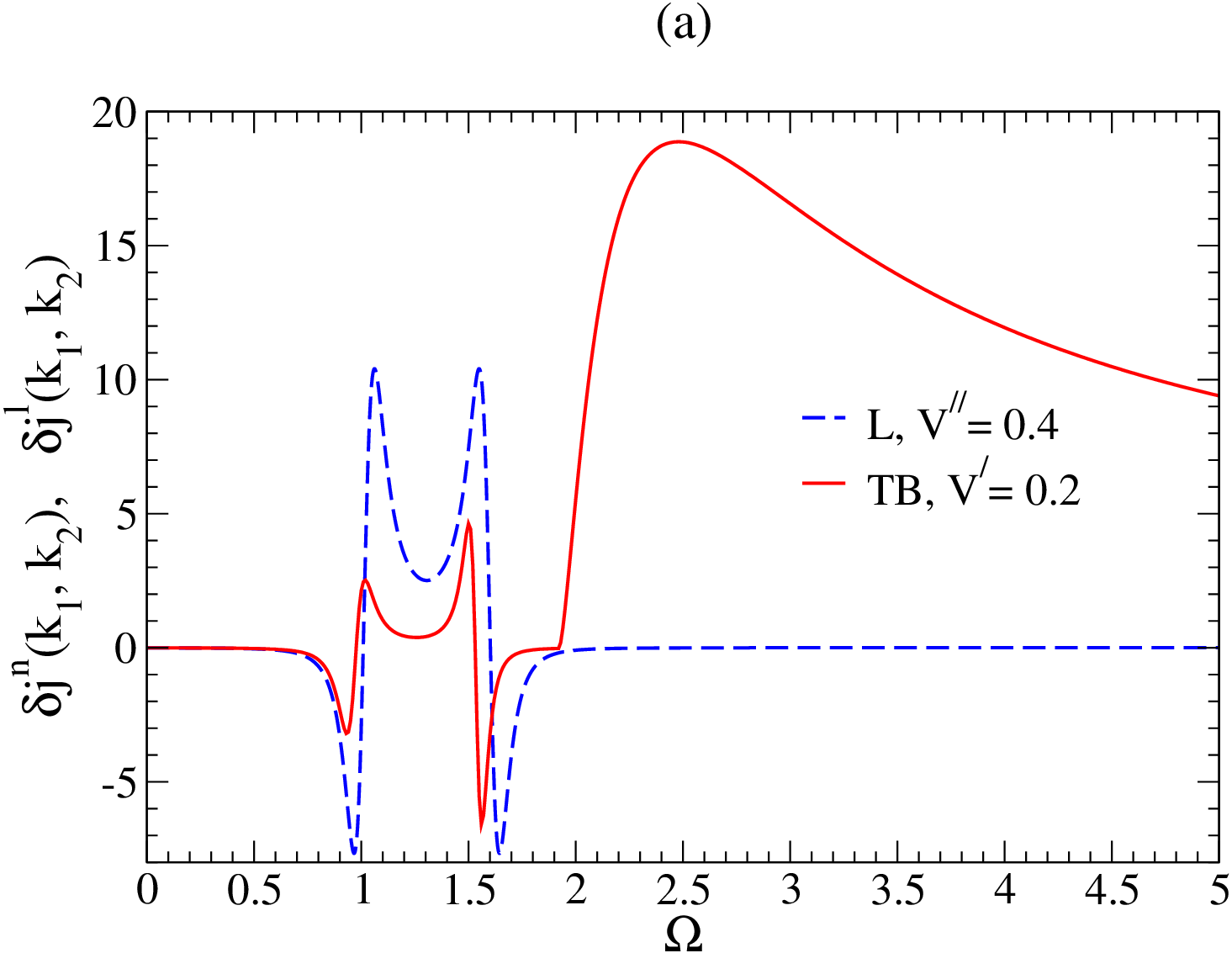,width=0.3\linewidth,clip=} &
\epsfig{file=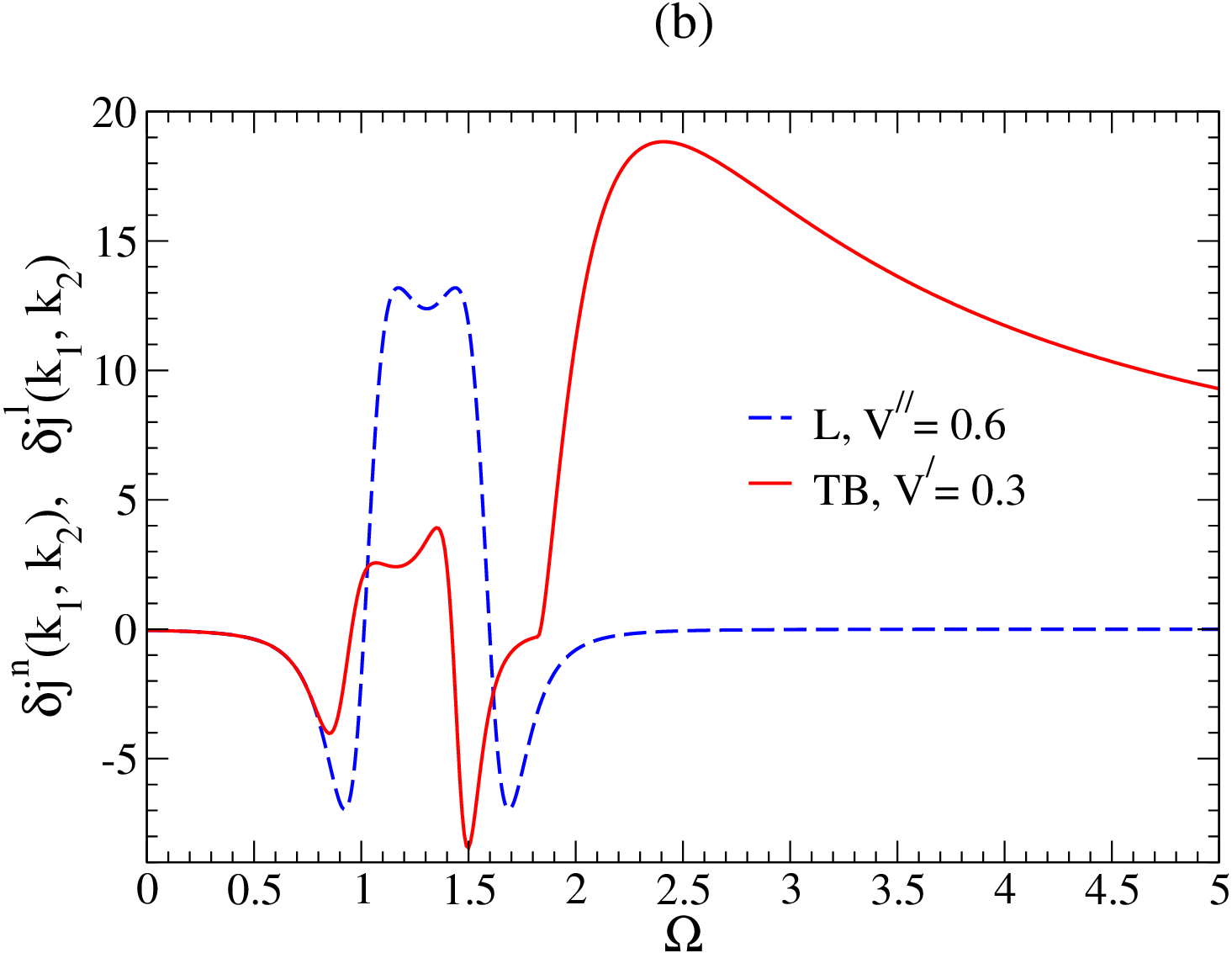,width=0.3\linewidth,clip=} &
\epsfig{file=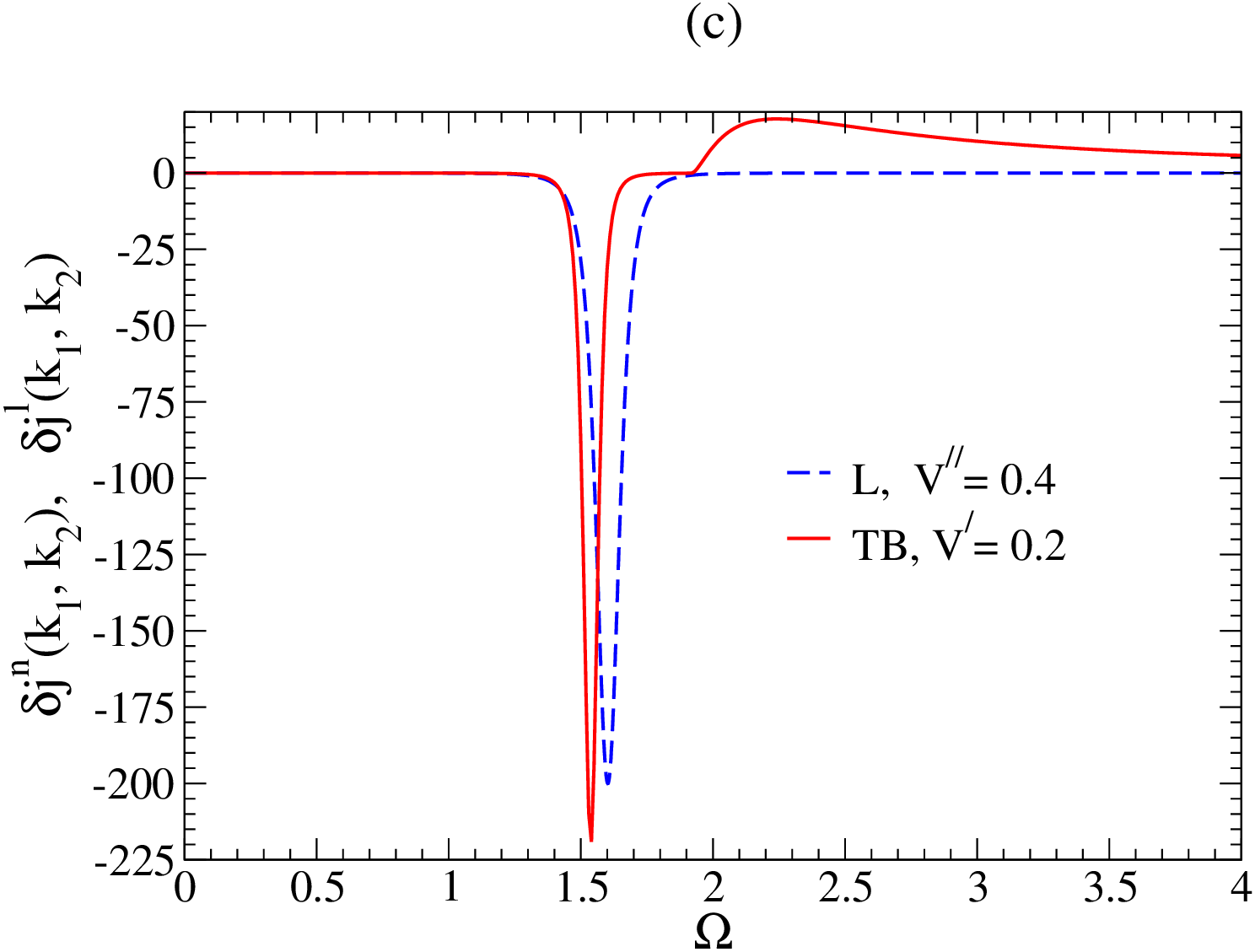,width=0.3\linewidth,clip=}
\end{tabular}
\caption{(Color online) Plot of the two-photon current (transmission) change due to interaction at the TLS site for a linear (L) and a tight-binding (TB) sinusoidal dispersion. The energies of the incident photons are as follows: (a) $E_{k_1}=1.01$, $E_{k_2}=1.6$, (b) $E_{k_1}=1.01$, $E_{k_2}=1.6$, and (c) $E_{k_1}=1.6$, $E_{k_2}=1.6$. $E_{k_1}$, $E_{k_2}$, $\Omega$, $V'$, and $V''$ are in units of $J$, with $J=1$ and $\hbar=1$.}
\label{pl4} 
\end{figure*}

The two-photon current (or transmission) has been calculated exactly by taking the expectation value of the current operator $\hat I_x$ in the scattering state $|\psi_{\bf k} \ra = |\phi_\bk \ra+| S_\bk \ra$, where $|S_\bk \ra \equiv G_0^+(E_{\bk}) \mathcal{V} |\psi_\bk \ra$ is the interaction-induced correction to the scattering state. The two-photon current $I^n(k_1,k_2)~(\equiv \la \psi_{\bk} |\hat I_x |\psi_{\bk} \ra)$ has two parts: one, $j^n_I(k_1,k_2)~(= \la \phi_\bk | \hat I_x |\phi_\bk \ra)$, is the contribution from two noninteracting photons, and the other, $\delta j^n(k_1,k_2)~(\equiv j_C +j_S)$, is induced  by the nonlinear photon-photon interaction at the impurity site. Thus, $I^n(k_1,k_2)=j^n_I(k_1,k_2)+\delta j^n(k_1,k_2)$. Here, $j_C~(= \la \phi_\bk | \hat I_x | S_\bk \ra + \la S_\bk | \hat I_x | \phi_\bk \ra)$ is an expectation value of the current operator between scattered and incident photon wave functions, and it is a measure of two-photon cross correlation. An expectation value of $\hat I_x$ in the interaction-induced scattered wave function is given by $j_S = \la S_\bk | \hat I_x | S_\bk \ra$. We now determine all these terms separately. We find that $j^n_I(k_1,k_2) = {\cal N}[I^n(k_1) + I^n(k_2)]$, where $\cal N$ is the total number of sites in the system. 
 For $x>1$ or $x<-1$, we find $j_S = 2J~{\rm Im} ~\la S_{\bf k}|a^\dg_x a_{x+1}|S_{\bf k}\ra$ with
\bea 
\la S_\bk|a^\dg_x a_{x+1}|S_\bk\ra &=& \f{2|\phi_\bk( {\b0})|^2}{|K_{
E_\bk}({\b0})|^2} \int_{-\pi}^\pi \f{dq}{2\pi} |\phi_q(0)|^2I_0(q) I_1^*(q), \nn \\
{\rm where}~I_m (q) &=& \int_{-\pi}^{\pi} \f{dq_1}{2\pi} 
\f{\phi_{q_1}(0) \phi^*_{q_1}(x+m)}{E_\bk -E_{qq_1} -i\n},~m=0,1,\nn
 \eea
and, $j_C = 2J~ {\rm Im} ~\la \phi_\bk | (a^\dg_x a_{x+1} - a^\dg_{x+1} a_x) |S_\bk \ra$ with
\bea & & \la \phi_\bk| a^\dg_{x_1} a_{x_2} |S_\bk\ra=- \f{\phi_\bk ({\b0})}{
K_{E_\bk}({\b0})}\times \nn \\ & & ~\int^\pi_{-\pi} \f{dq}{2\pi} ~\phi_q (x_2) \Big( \f{\phi^*_{k_2}(x_1) \phi^*_{k_1 q}({\b0})}{E_{k_2} -E_q
+ i\n} + \f{\phi^*_{k_1}(x_1) \phi^*_{k_2 q}({\b0})}{E_{k_1}- E_q +i\n} \Big).\nn
\eea

\begin{figure}
\includegraphics[width=7.0cm]{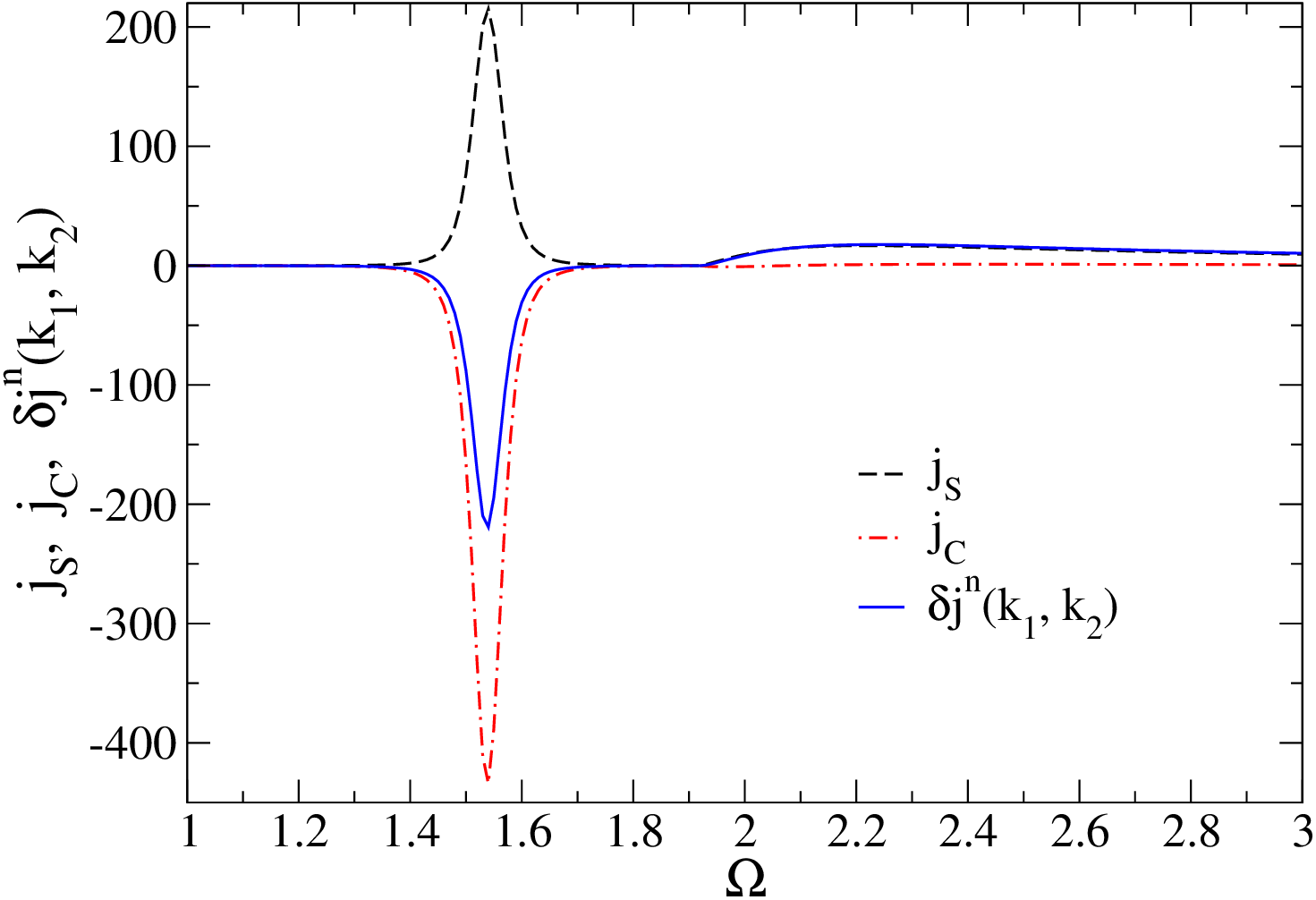}
\caption{(Color online) Two-photon current change $j_S$, $j_C$, and $\delta j^n(k_1,k_2)$ at $x>0$ for a tight-binding sinusoidal dispersion with energy of the incident photons $E_{k_1}=E_{k_2}=1.6$ and $V'=0.2$. $E_{k_1}$, $E_{k_2}$, $\Omega$ and $V'$ are in units of $J$, with $J=1$ and $\hbar=1$.}
\label{pl5} 
\end{figure}

Thus, $j^n_I(k_1,k_2)$ is one order of magnitude higher than $\delta j^n(k_1,k_2)$ for a finite system \cite{Dhar08}.  In the redefined system described by the Hamiltonian in Eq. (\ref{Ham2}), the two photons see each other only when both are at the impurity site, i.e., at $x=0$, and the amplitude for this to occur is  on the order of $1/\mathcal{N}$ for a finite system of $\mathcal{N}$ sites. In Fig.\ref{pl4}, we plot the interaction-induced two-photon current change $\delta j^n(k_1,k_2)$ for different values of $V'$ and energy of the incident photons.  The current change $\delta j^n(k_1,k_2)$ is nonzero near the single-photon resonances. We find that the direct correlation $j_S$ has the same sign (which is positive for transmitted photons) for all values of $\Omega$, whereas the cross correlation $j_C$ can have different signs depending on $\Omega$.  A two-photon bound state results in the two-photon scattering state for both  the linear \cite{Shen07} and the nonlinear dispersions \cite{Dhar08, Roy10b}. The bound state acts as a composite object and remains together after passing through the TLS. It has a maximum contribution in the two-particle current change for $E_{k_1}=E_{k_2}=\hbar\Omega$, which is manifested by $j_S$ in Fig. \ref{pl5}. The energy and momentum of the scattered photons are redistributed over a wide range of values satisfying the total energy conservation \cite{note1}. This redistribution of energy and momentum  emerges as the background fluorescence \cite{Shen07}, which can be conceived as being a result of the inelastic scattering of one photon from a composite transient object formed by the TLS and the other photon \cite{note2}. The magnitude of $\delta j^n(k_1,k_2)$ around the single-photon resonance is quite large when the energy of the incident photons is the same [see Fig. \ref{pl4}(c)]. We find in Fig.\ref{pl5} that the cross correlation $j_C$ has a large negative value near the single-photon resonance. A photon spends more time near the resonance and thus the effective strength of photon-photon interaction is increased much when both the incident photons have the same energy. One can also interpret the large reduction of $\delta j^n(k_1,k_2)$ near the single-photon resonance as being due to  antibunching of the scattered photons \cite{Shen07} or as a manifestation of the photon blockade. 
 
Surprisingly, we find a large enhancement of $\delta j^n(k_1,k_2)$ for a certain value of the transition energy $\hbar \Omega$, which is away from any of the single-photon resonances. It occurs at the value of $\hbar \Omega$ where $K_{E_{\bk}}(\b0)$ becomes zero, and it is relatively more pronounced for the incident photons with different energies as shown in Figs. \ref{pl4}(a) and \ref{pl4}(b). This is a special ``two-photon resonance'' arising for a finite bandwidth of the dispersion \cite{Roy09, Economou06}. The direct correlation term $j_S$ only contributes in $\delta j^n(k_1,k_2)$ at this special two-photon resonance. We can also phrase that a special ``two-photon bound state'' is formed for the finite bandwidth TB dispersion, and it creates a strong bunching of the transmitted photons. Interestingly, the magnitude of $\delta j^n(k_1,k_2)$ at  the two-photon resonance depends on the position of the nearest single-photon resonance. 

{\it Linear dispersion:} Now we determine the two-photon dynamics of $\mathcal{H}_l$. It was recently studied \cite{Shen07, Roy10} using the Bethe ansatz approach for a linearized dispersion of the Hamiltonian $\mathcal{H}_n$. We here construct the two-photon scattering eigenstates for $\mathcal{H}_l$ following Ref. \cite{Roy10} for an arbitrary $U$, and we show that our results in the limit $U\to \infty$ indeed match with the earlier results of Refs. \cite{Shen07, Roy10}. The general two-photon scattering eigenstate is of the form:
\begin{widetext}
\bea
&&\int dx_1dx_2\Big[A_2\big\{g(x_1,x_2)\f{1}{\sqrt{2}}c^{\dg}_e(x_1)c^{\dg}_e(x_2)+e(x_1)\delta(x_2)c^{\dg}_e(x_1)b^{\dg}+w~\delta(x_1)\delta(x_2)\f{1}{\sqrt{2}}b^{\dg}b^{\dg}\big\}+B_2\big\{t(x_1;x_2)c^{\dg}_e(x_1)c^{\dg}_o(x_2)\nn \\ &&+f(x_1)\delta(x_2)c^{\dg}_o(x_1)b^{\dg}\big\}+C_2h(x_1,x_2)\f{1}{\sqrt{2}}c^{\dg}_o(x_1)c^{\dg}_o(x_2)\Big]|0,0\ra 
\label{wavefn}
\eea 
\end{widetext}
where $g(x_1,x_2)$ and $h(x_1,x_2)$ are the probability amplitudes for both the photons in the even or the odd subspace, respectively, while $t(x_1,x_2)$ is the amplitude for one photon in the even and the other in the odd subspace. The quantity $e(x)~(f(x))$ is the probability amplitude for one photon in the $e~(o)$ subspace when the impurity is occupied by another photon; $w$ is the probability amplitude for both the photons at the impurity site. Here, $A_2,B_2$ and $C_2$ keep track of the incident photons.  We determine the amplitudes in Eq. (\ref{wavefn}) using the two-photon Schr{\"o}dinger equation.  The results are  given in Appendix \ref{app2}. We find, as $U \to \infty$, $w \to 0$, and the other amplitudes of the scattering state in Eq. (\ref{wavefn}) reduces to that of the TLS-waveguide obtained in Ref. \cite{Roy10} with the original Dicke-type Hamiltonian. Thus, we find an indirect proof for the validity of the mapping scheme for the TLS-waveguides on the Anderson impurity model with infinite $U$ in the present nonequilibrium dynamics study. 

 Here we estimate the two-photon current by taking the expectation value of $\hat{I}$ between the state in Eq. (\ref{wavefn}) in the limit $U \to \infty$. We choose  both the incoming photons from the left of the TLS; i.e., $A_2=V_1^2/V^2,B_2=V_1V_2/V^2$, and $C_2=V_2^2/V^2$. We find $I^l(k_1,k_2)=\mathcal{L}[I^l(k_1)+I^l(k_2)]/2\pi+\delta j^l(k_1,k_2)$, where $\mathcal{L}$ is the length of the full system, and 
\bea
\delta j^l(k_1,k_2)=\f{V_1^4V_2^2}{\pi^2 V^5} {\rm Im}\Big[e_{k_2}^2e_{k_1}^*+e_{k_1}^2e_{k_2}^*\Big],
\label{twocurr1}
\eea
where $\delta j^l(k_1,k_2)$ is the two-photon current change due to photon-photon interaction created by the TLS for the linear dispersion. We plot $\delta j^l(k_1,k_2)$ (after multiplying by $4\pi^2$) in Fig.\ref{pl4} with the transition energy $\hbar\Omega$ for different coupling strengths and incident energies. We find that $\delta j^l(k_1,k_2)$ is qualitatively similar to $\delta j^n(k_1,k_2)$ near the single-photon resonances. We again notice a large reduction of the two-photon current near the single-photon resonance when incident energy of the two photons is the same [see Fig.\ref{pl4}(c)]. The current change $\delta j^l(E_{k_1}=E_{k_2}=\hbar \Omega)=-16 V_1^4V_2^2v_g^3/(\pi^2V^8)$ is independent of the incoming photon's energy. The physical mechanism of the large reduction of $\delta j^l(k_1,k_2)$ near the single-photon resonance is similar to that of the TB dispersions. 
However, $\delta j^l(k_1,k_2)$ is almost zero near $\hbar\Omega \sim 2J$ for the linear dispersion. Thus there is no special two-photon resonance away from the single-photon resonances for the linear dispersion without band edges. It is quite different from the two-photon current line shape for the TB sinusoidal dispersion. The quantitative distinction between $\delta j^l(k_1,k_2)$ and  $\delta j^n(k_1,k_2)$ near the single-photon resonances can be attributed again to the variance in the group velocity of a photon for the linear and TB sinusoidal dispersions.

\section{Concluding remarks}
 In summary, we have shown qualitative and quantitative differences in the correlated two-photon transport in  1D waveguides for  linear and  TB sinusoidal dispersions.  A special ``two-photon resonance'' in the two-photon transmission line shape of the sinusoidal dispersion has been found. It might be useful to achieve a strong photon-photon interaction for nonmonochromatic light at low intensity. There have been enormous advances in experiments to realize optical nonlinearities at the few-photon level by creating strong light-matter interactions in various systems \cite{Akimov07,Astafiev10, Fushman08}. We hope that our prediction of two-photon resonance for a TB sinusoidal dispersion will be observed in the near future. We also plan to study the correlated dynamics of few photons in a cavity quantum electrodynamics setup with a two-level or  multilevel (driven) atom (or atoms) and photons with tight-binding sinusoidal dispersion, such as, in one-dimensional coupled-resonator-optical waveguides.
\section{acknowledgments}
The author would like to thank C. J. Bolech and S. Bose for fruitful discussions.

\appendix
\section{Derivation of the state in Eq. (\ref{2phscatts})}
\label{app1} 
The full scattering eigenstate of the Hamiltonian $\mathcal{H} = \mathcal{H}_0
+\mathcal{V}$ is given by the Lippman-Schwinger equation \bea |\psi\ra = |\phi\ra + G_0^+(E_{\bf k})\mathcal{V} |\psi\ra, \label{b1}\eea where the noninteracting two-photon Green's function is $G_0^+(E_{\bf k}) = 1/(E_{\bf k}- \mathcal{H}_0 +i \n)$. In the position basis $|\bx \ra~(\equiv \f{1}{\sqrt{2}}a^{\dg}_{x_1}a^{\dg}_{x_2}|\varnothing \ra,~{\rm where}~|\varnothing \ra$ is a vacuum), we obtain \bea \psi_\bk (\bx) ~=~ \phi_\bk (\bx) ~+~ UK_{E_\bk} (\bx)~ \psi_\bk (\b0), \label{b2} \eea
where $\psi_\bk (\b0) = \phi_\bk (\b0)/[1 - U K_{E_\bk} (\b0)]$ is derived from Eq. (\ref{b2}) by inserting ${\bf x}={\bf 0}$.
\section{Amplitudes of the state in Eq. (\ref{wavefn})}
\label{app2} The amplitudes in Eq.\ref{wavefn} are as follows:
\bea
&&g(x_1,x_2)=\f{1}{2\pi\sqrt{2}}g_{k_1}(x_1)g_{k_2}(x_2)-\f{iV}{\sqrt{2}v_g}C'e^{iE_{\bk}x_1/v_g}\nn \\&&\times~ e^{-i(\hbar\Omega-iV^2/2v_g)(x_1-x_2)/v_g}\theta(x_1-x_2)\theta(x_2)+(1 \leftrightarrow 2),\nn\\
&&~~~~~~e(x)=\f{1}{2\pi}\Big( g_{k_1}(x)e_{k_2}+g_{k_2}(x)e_{k_1}\Big)\nn \\&& ~~~~~~~~~~~~~~+C'e^{i(E_{\bk}-\hbar\Omega+iV^2/2v_g)x/v_g}\theta(x), \nn \\
&&t(x_1;x_2)=( g_{k_1}(x_1)h_{k_2}(x_2)+g_{k_2}(x_1)h_{k_1}(x_2))/2\pi,\nn \eea and 
$f(x)=( e_{k_1}h_{k_2}(x)+e_{k_2}h_{k_1}(x))/2\pi$,
$h(x_1,x_2)=(h_{k_1}(x_1)h_{k_2}(x_2)+h_{k_2}(x_1)h_{k_1}(x_2))/2\pi\sqrt{2}$, where $C'=(iV/\pi v_g)e_{k_1}e_{k_2}+\beta (e_{k_1}+e_{k_2})/(\pi(1-\beta))$ with $\beta=(iV^2/2v_g)/(\hbar\Omega+U/2-E_{\bk}/2)$.

$^\dagger$ Present address: Department of Physics, University of Cincinnati, Cincinnati, Ohio 45221, USA


\begin{thebibliography}{34}
\bibitem{note1} The total momentum of the two incident photons is not conserved after scattering for the TB dispersion, whereas the total momentum is automatically conserved by the energy conservation constraint for the linear dispersion.
\bibitem{note2} The redistribution in energy of the scattered photons is more explicit in the bound-state contributions (see, second term of $g(x_1,x_2)$ in Appendix \ref{app2}) of the state in Eq. (\ref{wavefn}) for the linear dispersion.
\end{thebibliography}
\end{document}